\documentclass[10pt,conference]{IEEEtran}
\usepackage{algpseudocode} 
\usepackage{subfig}
\usepackage{bm}
\usepackage{mathrsfs}
\usepackage{textcomp}
\usepackage{epsfig}
\usepackage{graphicx}
\usepackage{epstopdf}
\usepackage{indentfirst}
\usepackage{amsmath}
\usepackage{amssymb}
\usepackage{array}
\usepackage{multirow}
\usepackage{graphicx}
\usepackage{color, soul}
\usepackage{amsmath}
\usepackage{amsthm}

\usepackage{times}
\usepackage[ruled,vlined]{algorithm2e}
\theoremstyle{definition}

\theoremstyle{remark}

\usepackage{cite}

\def\BibTeX{{\rm B\kern-.05em{\sc i\kern-.025em b}\kern-.08em T\kern-.1667em\lower.7ex\hbox{E}\kern-.125emX}}
\usepackage{xcolor}

\usepackage{graphicx}
\usepackage{acronym}
\usepackage[T1]{fontenc}
\usepackage{aecompl}
\acrodef{BSS}[BSS]{Basic Service Set}
\acrodef{OBSS}[OBSS]{Overlaping Basic Service Set}
\acrodef{AP}[AP]{Access Point}
\acrodef{MAPC}[MAPC]{Multiple Access Point Coordination}
\acrodef{SNR}[SNR]{Signal to Noise Ratio}
\acrodef{MCS}[MCS]{Modulation and Coding Scheme}
\acrodef{QoS}[QoS]{Quality of Service}
\acrodef{MAB}[MAB]{Multi-Armed Bandit}
\acrodef{SNR}[SNR]{Signal to Noise Ratio}
\acrodef{SINR}[SINR]{Signal to Interference Plus Noise Ratio}
\acrodef{RSSI}[RSSI]{Received Signal Strength}
\acrodef{SR}[SR]{Spatial Reuse}
\acrodef{EDCA}[EDCA]{Enhanced Distributed Channel Access}
\acrodef{C-SR}[C-SR]{Coordinated Spatial Reuse}
\acrodef{C-OFDMA}[C-OFDMA]{Coordinated Orthogonal Frequency Division Multiple Access}
\acrodef{TXOP}[TXOP]{Transmission Opportunity} 
\acrodef{PPP}[PPP]{Poisson Point Process}
\acrodef{STA}[STA]{Station}
\acrodef{WLAN}[WLAN]{Wireless Local-Area Networks}
\acrodef{RL}[RL]{Reinforcement Learning}
\acrodef{MDP}[MDP]{Markov Decision Process}
\acrodef{RU}[RU]{Resource Unit}
\acrodef{TDD}[TDD]{Time Division Duplex}
\acrodef{DCF}[DCF]{Distributed Coordination Function}
\acrodef{ML}[ML]{Machine Learning}

\begin{document}

\title{Hierarchical Reinforcement Learning for Next Generation of Multi-AP Coordinated Spatial Reuse}

\thanks{
This work is supported in part by the Natural Science Foundation of Shaanxi Province of China(Program No.S2025-JC-QN-0514)}

 \author{
    \IEEEauthorblockN{Ziru Chen\IEEEauthorrefmark{1}, Salvatore Talarico\IEEEauthorrefmark{2}, Qing Xia\IEEEauthorrefmark{2},
    Xihan Peng\IEEEauthorrefmark{5},
     Xing Hao\IEEEauthorrefmark{6}, 
    Lin X. Cai\IEEEauthorrefmark{1}}

           \IEEEauthorblockA{\IEEEauthorrefmark{1}Department of Electrical and Computer Engineering, Illinois Institute of Technology, Chicago, USA}

     \IEEEauthorblockA{\IEEEauthorrefmark{2}Nokia Technologies, Wi-fi Center of Excellence, 520 Almanor Ave, Sunnyvale, CA 94085, USA}

     \IEEEauthorblockA{\IEEEauthorrefmark{5}Roku Inc, 1173 Coleman Ave, San Jose, CA 95110, USA}
     
     \IEEEauthorblockA{\IEEEauthorrefmark{6}School of Information Science and Technology, Northwest University, Xi’an, China}

}


\maketitle

\begin{abstract}

In next generation of Wi-Fi networks \ac{MAPC}  is poised to significantly enhance the network performance by enabling a set of \acp{AP} to coordinate with each other through advanced coordinating schemes so that to reduce
inter-AP contention and congestion. 
This paper focuses on defining a framework to facilitate the coordination across multi-APs when these employ \ac{C-SR}. In this case, the coordinating \acp{AP} may need to reciprocally adjust their scheduling strategy, power control and link adaptation to meet specific \ac{QoS} requirements, which by using classical approaches leads to high overhead due to negotiations needed across \acp{AP}, and requires complex solutions in order to properly optimize the network across all the parameters in play. In this matter,  
a two-layer \ac{MAB} algorithm has been proposed to optimize such a network while preserving the fair use of resources across all nodes.
The validity of this holistic approach is confirmed by system level simulations, which show that the proposed algorithm not only improves the network in terms of sum-throughput, but also allows to enhance fairness, making this a robust solution for next-generation of Wi-Fi networks.

\end{abstract}


\section{Introduction}

Over the past two decades Wi-Fi technology \cite{Ennis2023,Perahia2013} has consistently advanced to meet the needs of various applications by integrating state-of-the-art technologies and features to provide progressively higher transmission capacities and lower latency.
Beyond traditional Internet and video applications, Wi-Fi networks are nowadays expected to seamlessly integrate with the Internet of Things (IoT), healthcare, and virtual/augmented reality (VR/AR) applications. 
This convergence introduces new challenges, such as accommodating novel transmission requirements, managing higher device densities, and ensuring optimal energy efficiency. To address these demands, 
a common strategy has been proposed to increase the density of \acp{AP} within a given area, allowing devices to achieve higher \ac{SNR} levels and, consequently, higher transmission rates. However, as the number of \acp{AP} in a given area increases, the limited availability of frequency channels can lead to heightened contention and interference, thereby compromising the reliability of the Wi-Fi services. 
This approach can also lead to an uneven share of the radio resources across all nodes and some non-\ac{AP} \acp{STA} could be posed in the condition to starve in terms of \acp{TXOP} which reduces the user experience and degrades the overall network performance.

Currently, the IEEE working group devoted to evolving specifications for \ac{WLAN}
is working toward the design of Wi-Fi 8 \cite{Giordano2024}, and one of the proposed solutions to mitigate high contention levels in dense Wi-Fi environments is to implement coordinated transmission strategies among multiple \acp{AP}, such as \ac{SR}~\cite{imputato2024beyond,nunez2023multi,Jetmir2024,10427467, nunez2022txop,aio2019coordinated,nunez2024spatial}. This type of \ac{MAPC} scheme allows a set of \acp{AP} to coordinate with each other and negotiate proper use of the radio resources, such as specific \acp{RU} for each associated non-\ac{AP} \acp{STA}, the \ac{AP}'s transmit power and the \ac{MCS} associated to each transmission link, so that to reduce inter-\ac{AP} contention and congestion, but also 
allow to meet certain \ac{QoS} requirements. Additionally, the association among \acp{AP} and non-\ac{AP} \acp{STA} can be managed so that to allow an homogeneous resouce allocation among all nodes so that preserve fairness in terms of allowing a proper load balancing, which we call here network fairness.Considering that this is a multi-dimensional problem, classical approaches may lead to high overhead and long convergence time due to negotiations needed across the coordinating \acp{AP}, and requires complex solutions in order to properly optimize the network across all the parameters in play. In this matter, one alternative is to employ tools from \ac{ML} \cite{kihira2022interference, TODO, han2020deep,bellouch2024dql} to design a framework to address the aforementioned challenges, which is the main scope of this paper. In particular, the key contribution of this manuscript consists in a novel \ac{MAPC} framework, which leverages over \ac{C-SR}, where a two layer \ac{MAB} algorithm is used to jointly tune the coordinating \acp{AP}' scheduling, and transmit power, and also select the proper \ac{MCS} so that to achieve specific \ac{QoS} requirements associated to each link. The network is optimized by introducing a new metric, which aims at maximizing the network data rate, while preserving network fairness.

\section{Related Work}\label{re}
This section provides to the best of the authors' knowledge a comprehensive review of the studies conducted on \ac{C-SR} for \ac{WLAN} and the use of \ac{ML} in this context. 
In~\cite{imputato2024beyond}, the authors introduce a \ac{C-SR} interference model and present a strategy for estimating groups of APs that can transmit successfully simultaneously and which leverages on exchanging information across \acp{AP}. In~\cite{nunez2023multi}, the authors focused on the problem of creating multi-\ac{AP} groups that can transmit simultaneously to leverage spatial reuse opportunities, and studies different scheduling algorithms to determine which groups will transmit at every MAPC transmission.
In \cite{Jetmir2024}, the authors proposed a centralized \ac{C-SR} algorithm that determines the transmit powers of the concurrent \ac{AP} transmitters based on calculated interference levels in the main receiver.
 In \cite{10427467,nunez2024spatial,nunez2022txop,aio2019coordinated} the authors have provided comprehensive system level evaluations for \ac{C-SR} in order to  demonstrate the potential benefits of this coordinating scheme. For instance, the work in \cite{10427467} highlights the advantages of using \ac{C-SR} when this is combined with proportional fair scheduling for interference mitigation. 
In \cite{nunez2022txop}, the authors evaluated two coordinated \ac{TXOP} sharing strategies, where one of which additionally employs \ac{C-SR}. The evaluations show that by additionally adopting \ac{C-SR}, a significant throughput gain is achievable. In \cite{aio2019coordinated}, the authors have investigated the performance of \ac{C-SR} by comparing various flavours of this coordinating scheme with legacy \ac{TDD} design where an \ac{AP} always transmits with maximum transmit power. In this work, results have shown that regardless of the flavour of the coordinating scheme, \ac{C-SR} always outperform legacy \ac{TDD}. Finally in~\cite{nunez2024spatial}, the authors evaluated the performance of \ac{C-SR} by considering an implementation of this scheme where channel access and inter-\ac{AP} communication are performed over-the-air using the \ac{DCF}. In this case, \ac{C-SR} is shown to achieve significant higher throughput compared to the baseline. 
The authors in~\cite{kihira2022interference} propose a method for identifying interference-free \acp{AP} and a technique for reducing the amount of shared information between coordinated devices using Q-learning. In~\cite{TODO}, the authors introduce a \ac{MAB} algorithm to determine the scheduling strategies in a network employing \ac{C-SR} with the aim to maximize the network's sum data rate. However, this approach does not guarantee network fairness, and thus the \ac{QoS} requirements. Moreover, this work does not fully explore the potential impact of power control and link adaptation. In this matter, to the best of our knowledge, no existing studies have simultaneously addressed both network fairness and the optimization of the network throughput, nor have they explored how to adjust transmit power and \ac{MCS} adjustment while considering the network's \ac{QoS} requirements.
\section{System Model}\label{mapc}
This section provides an illustration of the network topology, based on which a novel \ac{MAPC} framework is developed. After presenting the \ac{MAPC} model, the problem formulation and the related optimisation objective function is formulated. 
\subsection{Network Topology}
\begin{figure}
    \centering
    \includegraphics[width=1\linewidth]{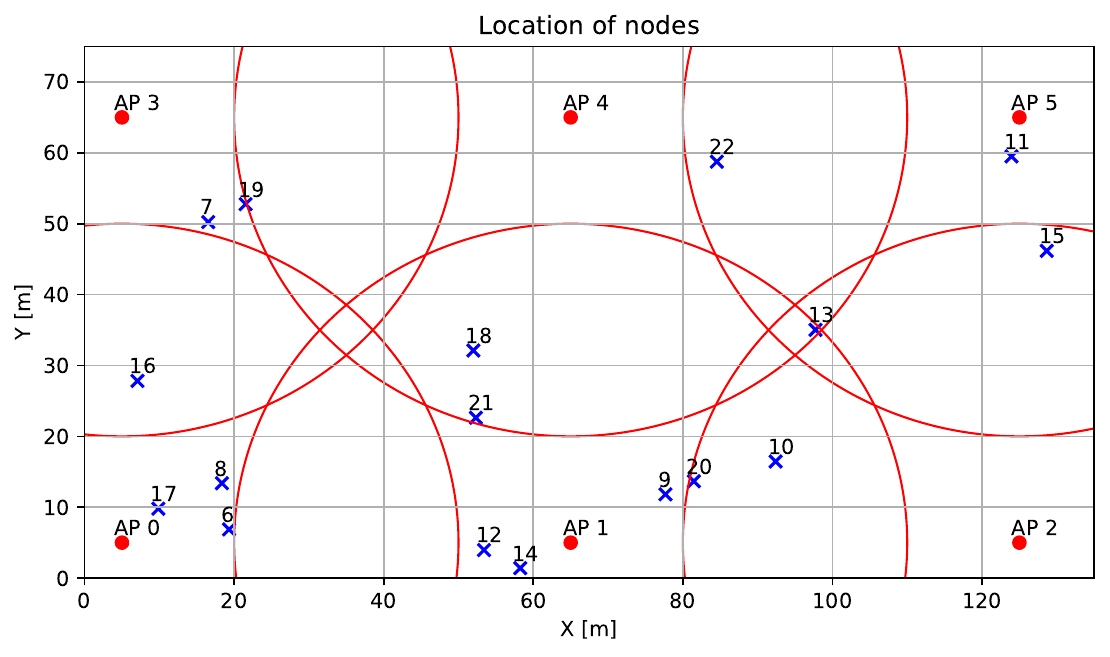}
    \caption{Illustration of an indoor office layout where the \acp{AP} are denoted by red dots and the non-\ac{AP} \acp{STA} are denoted by blue crosses.}
    \label{fig:topology}
\end{figure}

Consider a typical indoor office layout where $N=6$ \acp{BSS} are deployed. Assume that all BSSs have the same capability and coverage area when all the corresponding \acp{AP} transmit with the same transmit power. Each \ac{BSS} is modelled by a circular region with radius $r$ where the corresponding \ac{AP} is located in the center. Assume that \acp{AP} are deployed in a room of size 125 meters by 75 meters, and distributed so that each \ac{AP} is equally spaced with each other, and they are equally distanced from the perimeter walls of the room as shown in Fig. \ref{fig:topology}. Also let us assume that when \acp{AP} transmit using the maximum allowed power $r=45$ meters, and as illustrated in Fig. \ref{fig:topology} coverage areas of adjacent \acp{AP} overlap with each other. 
Without any loss of generality, all devices operates on the same frequency channel and same carrier bandwidth. Each \ac{AP} is associated with a given number of non-\ac{AP} \acp{STA}, which are randomly distributed in the whole area $A$ according to a \ac{PPP} with intensity $\lambda$, where the total number of deployed non-\ac{AP} \acp{STA} is denoted as $M$ and $\mathbb{E}[M] = \lambda|A|$.

In order to model the channel and penetration losses, the TGac NLOS path loss model for the residential case is used \cite{TGax2015}, where the path loss is given by: 
\begin{align}
    P_L = 40.05 + 20 \log_{10}{(\frac{\min(d,B_p)f_c}{2.4})} + P', \label{pathloss}
\end{align}
where $d$ represents the distance between devices in meters, $f_c$ is the central frequency in GHz, $B_p$ is the breaking point. Also when $d \geq B_p$, $P'$ is given by $P'=35log_{10}(d/B_p)$. Otherwise, $P'=0$.

For sake of simplicity, the assumption is to have an open floor with no walls traversed in both x-direction and y-direction, and it is assumed that each \ac{AP} is in full buffer with saturated traffic and unlimited size.

\subsection{MAPC Operation Framework }
After an initial discovery and pairing phase, which is outside the scope of this paper, and which is used to establish a coordinating set of \acp{AP}, one of the coordinating \acp{AP} serves as the transmission initiator, known as the sharing \ac{AP}, while the other coordinating \acp{AP} serve as followers and we refer to them as shared \acp{AP}. For simplicity in this paper, the selection of the sharing \ac{AP} among the coordinating set of \acp{AP} is done in a round-robin fashion.
Also,  it is assumed that each \ac{AP} may only serve a single non-\ac{AP} \ac{STA} per each \ac{TXOP}. The coordination among all coordinating \acp{AP} is done by a central controller that can be either a physical entity to which all \acp{AP} must be connected through a wireless or fiber backhaul or a logical entity collocated within any \ac{AP}, which is in charge of handling the coordination among \acp{AP} and the related resource management. 

\begin{figure}
    \centering
    \includegraphics[width=1\linewidth]{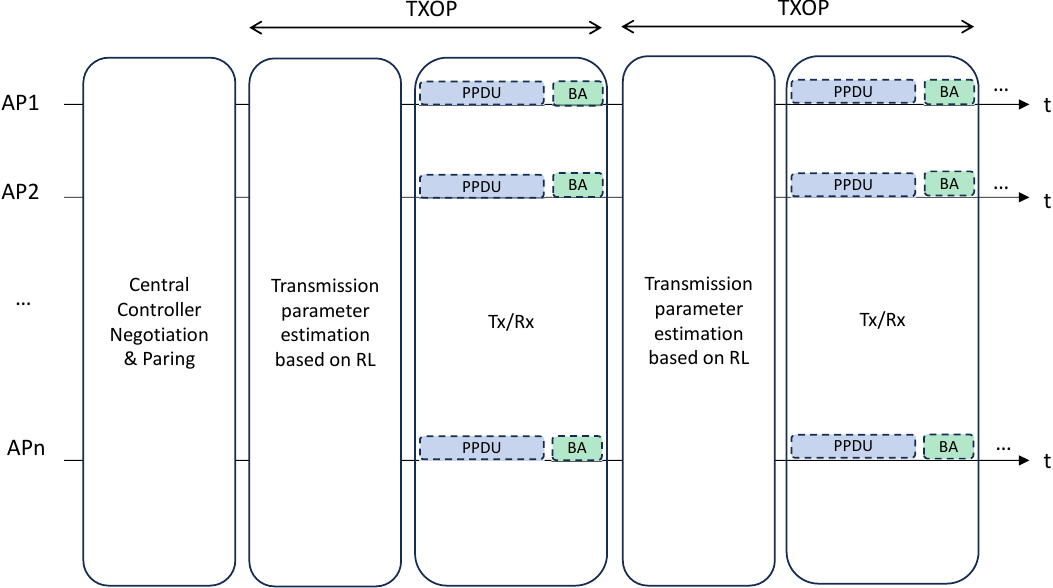}
    \caption{Illustration of the MAPC framework.}
    \label{fig:MAPC}
\end{figure}

In the rest of this paper, the index of the sharing \ac{AP} is denoted as $x$ and the associated non-\ac{AP} \ac{STA}, which is scheduled during a specific \ac{TXOP} is denoted as $y$. Once the sharing \ac{AP} and the scheduled non-\ac{AP} \ac{STA} among the associated non-\ac{AP} \acp{STA} are selected, the coordinating \acp{AP} will follow the instructions provided by the central controller in terms of scheduling and resource allocation. Specifically, the central controller indicates several control information to the coordinating set of \acp{AP}, which include sharing/shared \ac{AP} information, non-\ac{AP} \ac{STA} association, transmission power, \ac{MCS} selection, as well as \ac{QoS} requirements for all links and to all \acp{AP}. Once these instructions are communicated,  all \acp{AP} will be aware of their individual scheduling and resource management, and therefore can proceed transmitting downlink signals during a \ac{TXOP}. An illustration of the proposed \ac{MAPC} framework is summarized in Fig.~\ref{fig:MAPC}. 

In order to model power control in the $k$-th \ac{TXOP}, the set of transmit powers that can be associated to any \ac{AP} is denoted as $\mathcal{P}_k$, and the transmit power allocated to \ac{AP} $j$ during that instance of time is defined as $p_{j,k}\in \mathcal{P}_k$. It is important to note that the transmit power associated to any \ac{AP} is chosen from a discrete set $\mathbb{P}$ that is composed by $\mathbb{P} = [pl_{0}, pl_{1}, \cdots, pl_{d_t-1}, 0]$, where $pl_{z}$ is the $z$-th power level that is evaluated as $pl_{z} = \left(p_{max} - p_{min}\right)/{d_t}\cdot z + p_{min}$, $d_t$ is the number of power level that can be selected, and $p_{max}$ and $p_{min}$ are respectively the maximum and minimum transmit power that can be selected. Notice that $0$ is selected as the transmit power when an \ac{AP} is not scheduled for transmission during a specific \ac{TXOP}. As for link adaptation, the set of \acp{MCS} defined in the 802.11 specification for 20 MHz channel bandwidth \cite{10058126} are re-used assuming convolution codes are employed, and summarized in Table~\ref{MCS} 
. The set of \ac{MCS} indexes that can be selected during the $k$-th \ac{TXOP} is denoted as $\mathcal{M}_{k}$ where $m_{j,k}\in\mathcal{M}_{k}$ is the \ac{MCS} index for \ac{AP} $j$ at $k$-th \ac{TXOP}.   


\begin{table*}
\centering
\caption{MCS Table for 20MHz channels \cite{10058126}.}
\label{MCS}
\begin{center}
\begin{tabular}{ |c|c|c|c|c| }  
  \hline
 MCS index ($m_{j,k}$) & Modulation & Coding Rate & Data Rate (Mb/s), $\bar{R}(\cdot)$ & $\overline{SINR}$ (dB)\\
 \hline
  0 & BPSK & $1/2$ & $9$ & $10.61$ \\
  \hline
  1 & QPSK & $1/2$ & $17$ & $10.65$\\
  \hline
  2 & QPSK & $3/4$ & $26$ & $10.66$\\
  \hline
  3 & 16-QAM & $1/2$ & $34$ & $10.68$\\
  \hline
  4 & 16-QAM & $3/4$ & $52$ & $11.15$\\
  \hline
  5 & 64-QAM & $2/3$ & $69$ & $15.41$\\
  \hline
  6 & 64-QAM & $3/4$ & $77$ & $16.73$\\
  \hline
  7 & 64-QAM & $5/6$ & $86$ & $18.09$\\
  \hline
  8 & 256-QAM & $3/4$ & $103$ & $21.80$\\
  \hline
  9 & 256-QAM & $5/6$ & $115$ & $23.33$\\
  \hline
  10 & 1024-QAM & $3/4$ & $129$ & $29.78$\\
  \hline
  11 & 1024-QAM & $5/6$ & $143$ & $31.75$\\
  \hline
  12 & 4096-QAM & $3/4$ & $155$ & $33.74$\\
  \hline
  13 & 4096-QAM & $3/4$ & $172$ & $35.56$\\
  \hline
  14 & BPSK-DCM-DUP & $1/2$ & $N/A$ & $N/A$\\
  \hline
  15 & BPSK-DCM & $1/2$ & $4$ & $10.61$\\
  \hline
\end{tabular}
\end{center}
\end{table*}

Let us denote with $\mathcal{A}_{k}$ a set which indicates the active non-\ac{AP} \acp{STA} during the $k$-th \ac{TXOP} based on the association strategy, where  $a_{i,j,k}\in\mathcal{A}_{k}$ denotes whether $j$-th \ac{AP} and the $i$-th non-\ac{AP} \ac{STA} have an active link during the $k$-th \ac{TXOP}. Note that $a_{i,j,k}=1$ if a transmission occurs between $j$-th \ac{AP} and the $i$-th non-\ac{AP} \ac{STA} during the $k$-th \ac{TXOP}, and  $a_{i,j,k}=0$ otherwise. 

Given the $k$-th \ac{TXOP} during which the $j$-th \ac{AP} transmits to the $i$-th non-\ac{AP} \ac{STA}, $N_{i,j,k}$ represents the number of frames that can be transmitted during that \ac{TXOP} with duration $\tau$ where the duration of each frame is indicated $L_{frame}$ and depends on the specific \ac{MCS} that is used for that link. With that said, the aggregated sum data rate during the $k$-th \ac{TXOP} for the $j$-th \ac{AP} could be written as 
\begin{align}
    R_{j,k} = \sum_{i=1}^{D} r_{i,j,k}. 
\end{align} 
where $D$ is the number of non-\ac{AP} \acp{STA} associated with the $j$-th \ac{AP} with an active link during the $k$-th \ac{TXOP}, and 
\begin{align}
    r_{i,j,k} = \frac{L_{frame}N_{i,j,k}}{\tau},
\end{align}
With that said, the number of frames transmitted during the  $k$-th \ac{TXOP} between the $j$-th \ac{AP} and the $i$-th non-\ac{AP} \ac{STA} is given by
\begin{align}
    N_{i,j,k} = \frac{\bar{R}(m_{j,k}) \tau}{L_{frame}}\mathbb{I}[SINR_{i,j,k}\geq \gamma]P^{success}_{i,j,k}, 
\end{align}
where $\bar{R}(m_{j,k})$ is the achievable rate when the \ac{MCS} index $m_{j,k}$ is used and the related values are those captured in Table~\ref{MCS}, $\gamma$ is a threshold, which determines based on the \ac{SINR} level, denoted by $SINR_{i,j,k}$, whether the transmitted frames could be detected, and $\mathbb{I}\left[x\right]$ is a function which assumes value 1 if x is true, otherwise is equal to zero. Furthermore, the \ac{SINR} level is expressed as
\begin{align}
SINR_{i,j,k} = \frac{p_{j,k} P_{L_{i,j,k}}a_{i,j,k} }{\sum_{u\neq i}\sum_{v\neq j}p_{v,k} P_{L_{u,v,k}}a_{u,v,k}+\sigma^2 },
\end{align}
where $P_{L_{i,j,k}}$ indicates the path loss between \ac{AP} $j$ and non-\ac{AP} \ac{STA} $i$ during the $k$-th \ac{TXOP}, and calculated as in (\ref{pathloss}), and $\sigma$ is the white Gaussian noise. Finally, $P^{success}_{i,j,k}$ is the probability 
that a frame transmitted specifically with \ac{MCS} index $m_{j,k}$ could be successfully decoded, and is 
calculated as 
\begin{align}
P^{success}_{i,j,k}= \mathbf{P}(SINR_{i,j,k}\geq X),
\end{align}
where $X\thicksim\mathcal{N}(\mathbb{E}[\overline{SINR}_{m_{j,k}}],\sigma_m^2)$, and $\overline{SINR}_{m_{j,k}}$ is the mean \ac{SINR} level required to decode a frame transmitted with \ac{MCS} index $m_{j,k}$ and the related values are those captured in Table~\ref{MCS}. 


\subsection{Problem Formulation}
When designing a wireless system, there is an existing trade-off between optimizing the sum data rate of the whole network and ensuring a fair resource allocation across all nodes so that resources are distributed in a homogeneous manner. In this matter, in order to jointly maximize both a suitable cumulative optimization objective function is necessary. A common approach is to combine these two separate KPIs into a weighted objective or use a fairness-aware metric like the proportional fairness criterion. 
\subsubsection{Weighted Sum}
The objective function is expressed as a weighted sum of the total data rate and a fairness metric, i.e., Jain’s fairness index. By changing the weights, it is possible to adjust the relative importance of each KPI.


By denoting with $\alpha\in [0,1]$ a weight parameter to balance between maximizing data rate and fairness, the objective function is defined as
\begin{align}
  \max_{\mathcal{P},\mathcal{A},\mathcal{M},Q} \quad &\alpha \cdot \sum_{j=1}^{N} \sum_{k=1}^{K}R_{j,k} + (1 - \alpha) \cdot f(\mathbf{R})\label{Weighted_sum_obj}  \\
  \text{subject to} \quad & r_{i,j,k} \geq Q, \quad \forall k \label{weighedsum}
\end{align}
where $f(\mathbf{R})$ is the Jain's fairness index of all \acp{AP} and is defined as 
\begin{align}
f(\mathbf{R})= \frac{\left( \sum_{j=1}^{N} \sum_{k=1}^{K}R_{j,k} \right)^2}{N \sum_{j=1}^{N} \sum_{k=1}^{K} R_{j,k}^2}.
\end{align} 

Notice that the above objective function is constrained so that to guarantee certain \ac{QoS} requirements for each transmission between an \ac{AP} and its associated non-\ac{AP} \ac{STA}, where the requirement is denoted as $Q$.

\subsubsection{Proportional Fairness}

Proportional fairness aims to balance between maximizing the total data rate and ensuring that the resource allocation is fair to all users by maximizing the sum of logarithms of the data rates.

The objective function for proportional fairness is give by:
\begin{align}
\max_{\mathcal{P},\mathcal{A},\mathcal{M},Q} \quad &\sum_{j=1}^{N} \log(\sum_{k=1}^{K}R_{j,k})\label{Proportional_obj} \\
\text{subject to} \quad & r_{y,x,k} \geq Q, \quad \forall k. \label{propfair}
\end{align}
This function inherently balances maximizing the sum of the data rates while giving diminishing returns to higher rates, which encourages more fair resource allocation among users.
Notice that once again the above objective function is constrained so that to guarantee certain \ac{QoS} requirements for each transmission between an \ac{AP} and its associated non-\ac{AP} \ac{STA}, where the requirement is denoted as $Q$.

\section{Proposed Algorithm}\label{algorithm}

In this study,  a two-layer \ac{RL} algorithm is introduced to tackle the problem at hand. The core of \ac{RL} is the \ac{MDP}, where the next state is determined solely by the current state and action, without any dependence on past states. An \ac{MDP} is characterized by a set of states $S$, actions $A$, a reward function $R$, and state transition probabilities $P_t$, and can be defined as:
\begin{equation}
  M = <S,A,P_t,R>.  
\end{equation}
The state space is determined by the network topology, which significantly impacts the scheduling decisions. The action space is composed by $\mathcal{P} = \{\mathcal{P}_k,\forall k\in [1, K]\}$, which is the set of the transmit powers, $\mathcal{M} = \{\mathcal{M}_k,\forall k\in [1, K]\}$, which is the set of the MCS selections, $\mathcal{A} = \{\mathcal{A}_k,\forall k\in [1, K]\}$, which is the set of association strategies for $K$ TXOP,
and $Q$, which is the set of \ac{QoS} requirements. Notice that the objective is to maximize the overall data rate while ensuring fairness, as defined by the reward functions in equations~\eqref{Weighted_sum_obj} and~\eqref{Proportional_obj}.

Since these reward functions consider long-term fairness, they need to be evaluated over multiple time slots $K$. Updating all parameters only once every $K$ time slots could result in slow adaptation to dynamic changes. To address this, we propose a hierarchical \ac{MAB} approach. The outer layer \ac{MAB} algorithm adjusts the target \ac{QoS} parameters $Q$ between the sharing AP and its associated non-\ac{AP} \ac{STA} every $T_{\text{outer}}$ time slots, balancing data rate maximization and fairness as~\eqref{Weighted_sum_obj} and~\eqref{Proportional_obj}. The inner layer MAB algorithm, operating at each single time slot, fine-tunes $\mathcal{P}$, $\mathcal{A}$, and $\mathcal{M}$ based on the updated \ac{QoS} targets.
In the outer layer, the agent determines the \ac{QoS} parameters $Q$, which are implemented in simulations over a duration of $T_{outer}$ time slots to optimize the reward. The performance results from these simulations are then fed back to the agent to inform and refine its subsequent decisions.

In the inner layer, for each time slot, a sharing AP and one of its associated non-\ac{AP} \acp{STA} are randomly selected for downlink transmission. The 1st-level agent then selects a subset of shared APs based on the context (chosen AP and non-\ac{AP} \ac{STA}). Following this, the 2nd-level agent selects, for each AP in the subset, an associated non-\ac{AP} \ac{STA} along with its transmission power and \ac{MCS}. These selected configurations are tested through simulations to maximize the data rate, and the resulting performance metrics are fed back to the agents to guide future decisions.

To encourage exploration during training, a noise term is added to each action, sampled from a Gaussian distribution to balance between trying new actions and using known strategies. In the execution phase, the model remains static without further updates or reward collection. The noise term can be set to zero, allowing the model to make decisions based solely on the current state in the MAP. If there are changes in the environment, training can be re-initiated to adjust the neural network at predefined intervals. 



    

\section{Simulation Results}\label{sim}
\begin{table} [t!]
\centering
\caption{Simulation Parameters.}
\label{parameters}
\begin{center}
\begin{tabular}{ |c|c| } 
 \hline
$ \mathbf{Parameter}$ & $\mathbf{Value}$ \\ 
  \hline
 $\lambda$ & $0.002$ \\
 \hline
    $r [meter]$ & $45$ \\
     \hline
  $\sigma$ & $\mathcal{N}(0,1)$ \\
 \hline
  $K$ & $5000$ \\
 \hline
 $d_t$ & $8$ \\
 \hline
 $p_{max}[dB]$ & $20$ \\ 
 \hline
 $p_{min}[dB]$ & $10$ \\ 
 \hline
 $B_p$[meter] & $3$ \\ 
 \hline
 $f_c[GHz]$ & $2.4$ \\ 
 \hline
 $L_{frame}[byte]$ & $1500$ \\ 
 \hline
 $\sigma_m^2$ & $2$ \\
 \hline
  $\alpha$ & $0.02 $ \\
 \hline
   $\tau [sec]$ & $5.484e^{-3} $ \\
 \hline
\end{tabular}
\end{center}
\end{table}

This section provides a comprehensive performance evaluation of the proposed algorithm when this is compared with the baseline algorithm proposed in~\cite{TODO}. A Python-based simulator was thoroughly developed to simulate the Wi-Fi network operation, which also includes providing network state information, executing corresponding actions, and tracking network throughput. The proposed algorithm is implemented using a PyTorch framework. The performance results of the proposed two layer \ac{MAB} algorithm which are illustrated in this section are obtained using the set of simulation parameters summarized in Table~\ref{parameters}.

\subsection{Convergence of the Two-Layer \ac{MAB} Algorithm}

\begin{figure}[t!]
    \centering
    \includegraphics[width=1\linewidth]{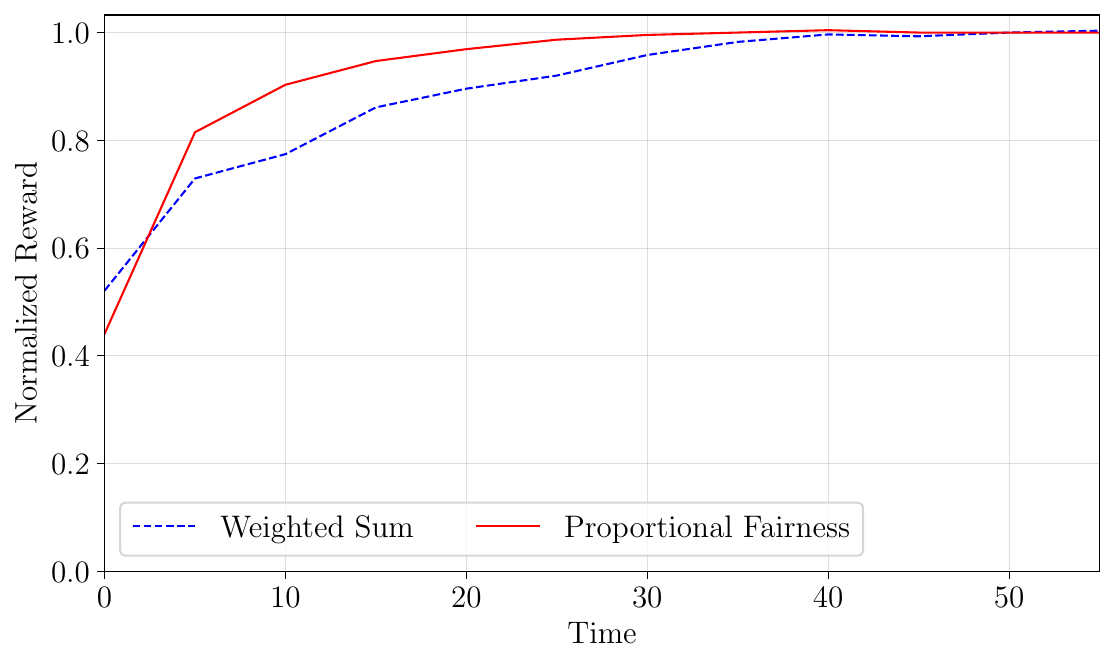}
    \caption{Comparison between weighted sum and proportional sum fairness reward in terms of convergence of the algorithm in function of time. }
    \label{fig:convergence}
\end{figure}

Fig.~\ref{fig:convergence} provides a comparison between the weighted sum (\ref{Weighted_sum_obj}) and proportional sum fairness (\ref{propfair}) reward in terms of convergence of the proposed algorithm. The figure indicates that the proposed two layer hierarchical MAB algorithm achieves convergence after few iterations. This rapid convergence demonstrates that the proposed algorithm is capable to quickly adapt to and optimize the network parameters, and simultaneously, to guarantee the reliable performance with respect to both rewards. In addition, it is also possible to notice that comparing with the weighted sum reward, the proportional sum reward is able to stabilize much faster. 

\subsection{Aggregated Sum Data Rate}

\begin{figure}[t!]
    \centering
    \includegraphics[width=1\linewidth]{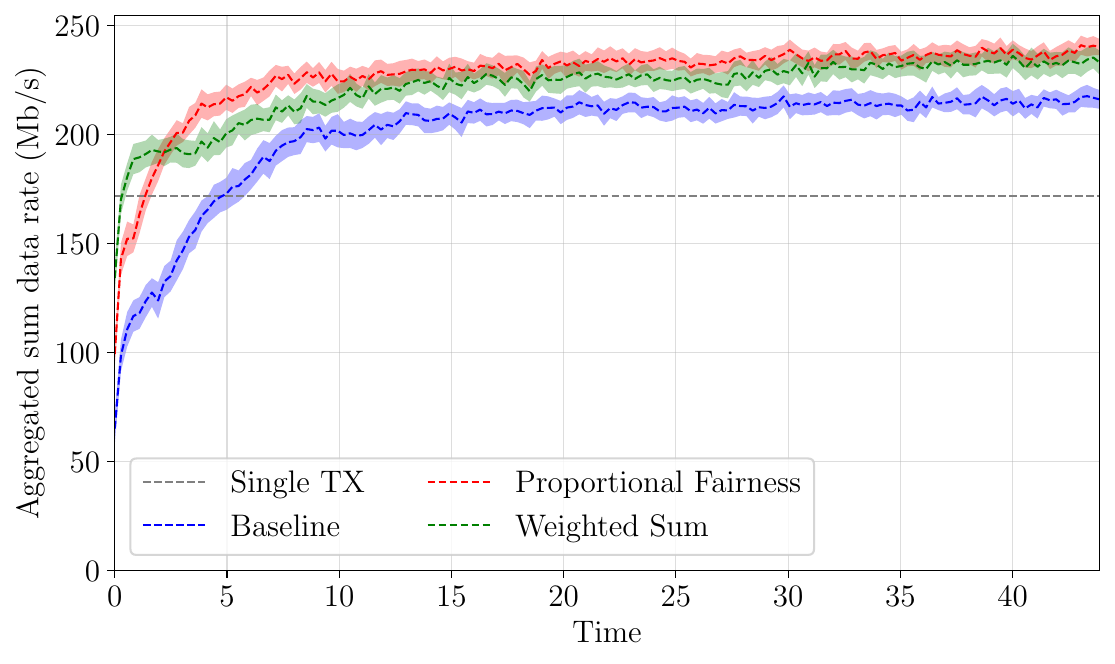}
    \caption{Aggregated sum data rate over all \acp{AP} in function of time.}
    \label{fig:sum_data_rate}
\end{figure}

Fig.~\ref{fig:sum_data_rate} illustrated the aggregated sum data rate over all \acp{AP} in function of time. The figure compares the performances between the case when a single \ac{AP} transmits with no \ac{MAPC}, with the case when \ac{MAPC} is used. In this case, the baseline algorithm is compared with our proposed solution when the two different rewarding strategies are used. Furthermore, for our proposed algorithm results are provided for each reward strategy introduced in Sec \ref{algorithm}. Fig.~\ref{fig:sum_data_rate} validates the effectiveness of adopting \ac{MAPC} when \ac{C-SR} is used, and shows that significant improvement can be achieved by employing such coordinating scheme. Fig.~\ref{fig:sum_data_rate} also shows that the proposed solution is able to achieve higher aggregated sum data rate compared to the baseline. 
Fig.~\ref{fig:sum_data_rate} also illustrates once again that in terms of reward strategy, the proportional fairness reward allows the algorithm to converge faster than the weighted sum reward, but also allows to achieve a slightly higher aggregated sum data rate.

\subsection{Network Fairness}
\begin{figure}[t!]
    \centering
    \includegraphics[width=1\linewidth]{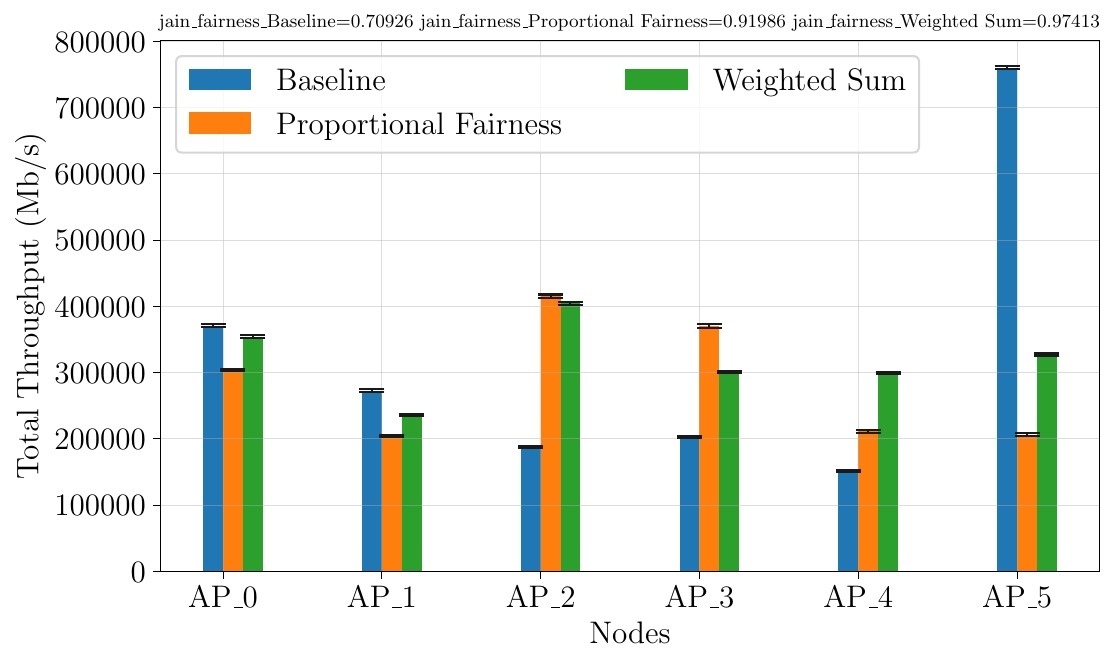}
    \caption{Average throughput evaluated for each individual AP.}
    \label{fig:fairness_ap}
\end{figure}

In order to show the benefits of the proposed algorithm in terms of network fairness, Fig. \ref{fig:fairness_ap} shows the average throughput evaluated for each individual AP and calculated over 
5000 \acp{TXOP}. This figure compares the baseline with the proposed algorithm when both the weighted sum (\ref{Weighted_sum_obj}) and proportional sum fairness (\ref{propfair}) reward are used. Fig. \ref{fig:fairness_ap} illustrates that while the baseline algorithm tends to favor the transmission of specific \acp{AP} (e.g., AP$\_$5 has much higher throughput than all other \acp{AP}), by using the proposed algorithm \acp{AP} will have a more balanced throughput, which is even more remarked when the weighted sum reward is used as the average throughput per \ac{AP} tends to be quite similar among all \acp{AP}. When using the Jain’s fairness index to quantify fairness, its value is equivalent to $0.709$ for the baseline algorithm, $0.92$ when our proposed algorithm is used in combination with the proportional fairness reward, and $0.97$ when our proposed algorithm is used in combination with the weighed sum reward.


\section{Conclusion}\label{con}
In this paper a novel framework based on a hierarchical \ac{RL} algorithm is proposed
with the aim to facilitate the coordination across multi-APs when these employ \ac{C-SR}.
The proposed framework provides an efficient solution to jointly tune the coordinating APs’ scheduling, and transmit power, but also select the proper MCS so that to
achieve specific \ac{QoS} requirements associated to each link. The network is optimized by introducing two new metrics, which aims at maximizing the network data rate, while ensuring the radio resources are always distributed homogeneously across all nodes so that to preserve network fairness.
A system-level simulation campaign has been performed to validate the effectiveness of the proposed solution, which highlights its ability to enhance the global throughput and network fairness. This innovative framework offers a promising and robust solution for optimizing resource allocation in the next-generation Wi-Fi networks, paving the way for improved user experiences and more reliable network performance in dense deployment scenarios.


\bibliographystyle{IEEEtran}
\bibliography{IEEEfull,ref}

\end{document}